\documentclass[aps,prl,twocolumn,superscriptaddress]{revtex4-1}
\usepackage{amsmath}	
\usepackage{graphicx}	
\usepackage{color}
\usepackage{gensymb}
\usepackage{braket}
\usepackage{hyperref}
\usepackage{upgreek}

\begin{document}
	
\title{Giant effective Zeeman splitting in a monolayer semiconductor realized by spin-selective strong light-matter coupling}
	
\author{T. P. Lyons}
\email{t.lyons@sheffield.ac.uk}
\author{D. J. Gillard}
\affiliation{Department of Physics and Astronomy, The University of Sheffield, Sheffield S3 7RH, UK}
\author{C. Leblanc}
\affiliation{Institut Pascal, PHOTON-N2, CNRS, Universit\'e Clermont Auvergne, F63000 Clermont-Ferrand, France}
\author{J. Puebla}
\affiliation{Center for Emergent Matter Science, RIKEN, Wako 351-0198, Japan}
\author{D. D. Solnyshkov}
\affiliation{Institut Pascal, PHOTON-N2, CNRS, Universit\'e Clermont Auvergne, F63000 Clermont-Ferrand, France}
\affiliation{Institut Universitaire de France (IUF), F-75231 Paris, France}
\author{L. Klompmaker}
\affiliation{Experimentelle Physik 2, Technische Universit\"at Dortmund, 44221 Dortmund, Germany}
\author{I. A. Akimov}
\affiliation{Experimentelle Physik 2, Technische Universit\"at Dortmund, 44221 Dortmund, Germany}
\affiliation{Ioffe Institute, Russian Academy of Sciences, 194021 St. Petersburg, Russia}
\author{C. Louca}
\affiliation{Department of Physics and Astronomy, The University of Sheffield, Sheffield S3 7RH, UK}
\author{P. Muduli}
\affiliation{Institute for Solid State Physics, University of Tokyo, Kashiwa 277-8581, Japan}
\affiliation{Department of Physics, Indian Institute of Technology Madras, Chennai 600036, India}
\author{A. Genco}
\affiliation{Department of Physics and Astronomy, The University of Sheffield, Sheffield S3 7RH, UK}
\author{M. Bayer}
\affiliation{Experimentelle Physik 2, Technische Universit\"at Dortmund, 44221 Dortmund, Germany}
\affiliation{Ioffe Institute, Russian Academy of Sciences, 194021 St. Petersburg, Russia}
\author{Y. Otani}
\affiliation{Center for Emergent Matter Science, RIKEN, Wako 351-0198, Japan}
\affiliation{Institute for Solid State Physics, University of Tokyo, Kashiwa 277-8581, Japan}
\author{G. Malpuech}
\affiliation{Institut Pascal, PHOTON-N2, CNRS, Universit\'e Clermont Auvergne, F63000 Clermont-Ferrand, France}
\author{A. I. Tartakovskii}
\email{a.tartakovskii@sheffield.ac.uk}
\affiliation{Department of Physics and Astronomy, The University of Sheffield, Sheffield S3 7RH, UK}
	
\date{\today}

\begin{abstract}
	Strong coupling between light and the fundamental excitations of a two-dimensional electron gas (2DEG) are of foundational importance both to pure physics and to the understanding and development of future photonic nanotechnologies \cite{smolka2014cavity,Efimkin2017,Back2017,Sidler2017,BingTan2020,Klein2009,Roch2019}. Here we study the relationship between spin polarization of a 2DEG in a monolayer semiconductor, MoSe$_2$, and light-matter interactions modified by a zero-dimensional optical microcavity. We find robust spin-susceptibility of the 2DEG to simultaneously enhance and suppress trion-polariton formation in opposite photon helicities. This leads to observation of a giant effective valley Zeeman splitting for trion-polaritons (g-factor $>20$), exceeding the purely trionic splitting by over five times. Going further, we observe robust effective optical non-linearity arising from the highly non-linear behavior of the valley-specific strong light-matter coupling regime, and allowing all-optical tuning of the polaritonic Zeeman splitting from 4 to $>10$ meV. Our experiments lay the groundwork for engineering quantum-Hall-like phases with true unidirectionality in monolayer semiconductors, accompanied by giant effective photonic non-linearities rooted in many-body exciton-electron correlations.
\end{abstract}
	
	
\maketitle

\section{Main}

\begin{figure*}
	\center
	\includegraphics[scale=1]{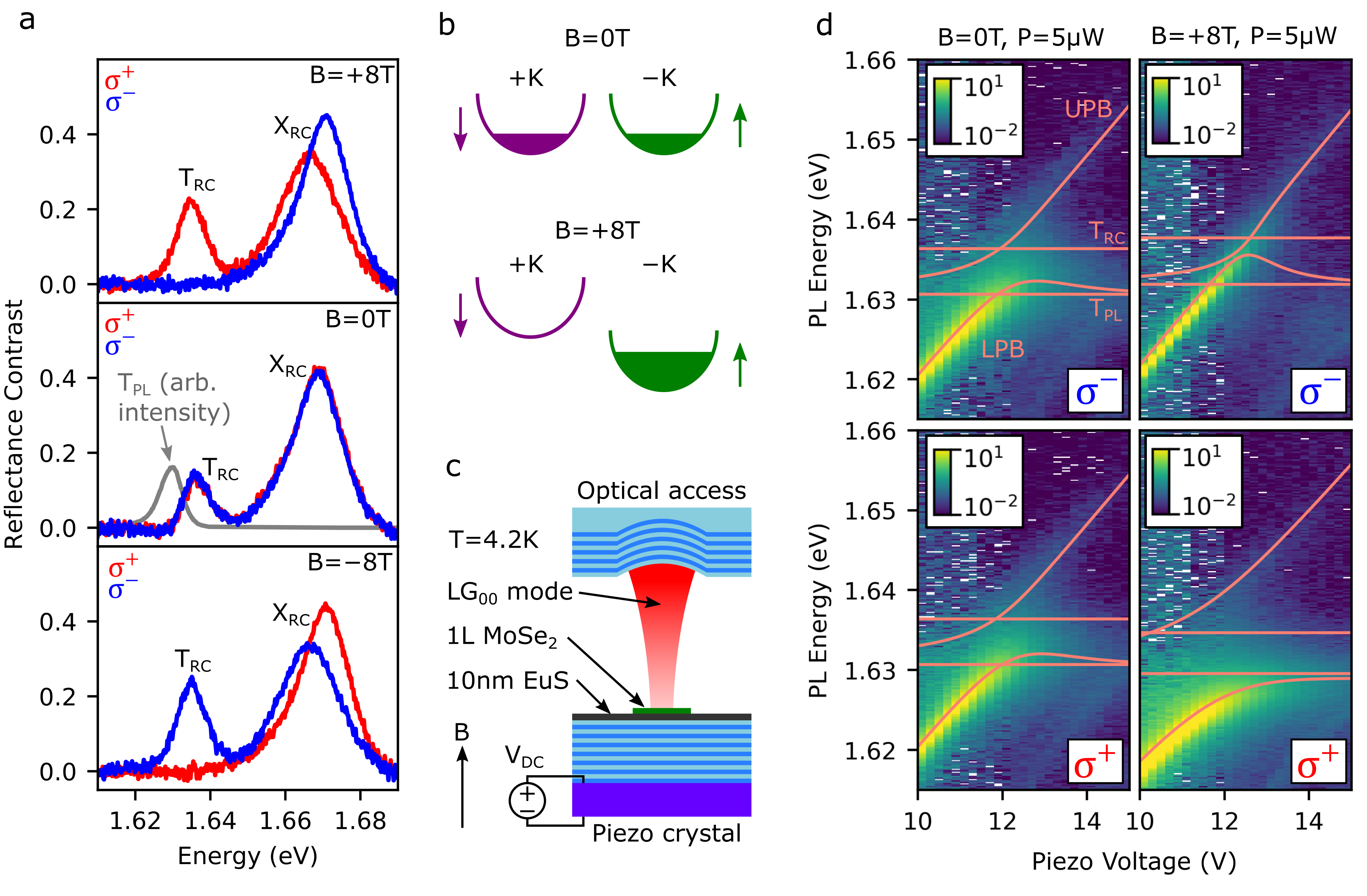}
	\caption{\label{fig:figure1} \textbf{Excitations of a 2-dimensional electron gas strongly coupled to light in monolayer MoSe$_2$.} (a) Reflectance contrast RC $=(R_0 - R) /R_0$ from monolayer MoSe$_2$ (reflectance $R$ on flake and $R_0$ on substrate) with raised itinerant carrier density at T $=4.2$ K and B $=-8,0,+8$ T. Two peaks are attributed to the neutral exciton ($X_{RC}$) and charged exciton or trion ($T_{RC}$). At high B-fields the trion absorption is completely suppressed in one or the other circular polarization of light. For comparison the trion photoluminescence $T_{PL}$ signal at $B=0$ T is also shown, revealing a Stokes shift of $\sim 6$ meV. (b) At B $=0$ T, the 2DEG has zero net spin polarization. At B $=+8$ T, the 2DEG is completely spin polarized, causing the oscillator strength of the $-K$ valley trion to be suppressed owing to a lack of itinerant electrons in the $+K$ valley. (c) Schematic of the open cavity structure used in this work. The cavity is formed by bringing a concave top DBR into the optical path above the planar bottom DBR, on top of which is the 10 nm EuS film and monolayer MoSe$_2$. The EuS film serves to increase the itinerant electron density in the MoSe$_2$. A vacuum gap separates the DBRs forming a zero-dimensional optical microcavity. Piezo nanopositioners allow precise tuning of the cavity length, whereby applying a DC voltage will decrease the cavity length and increase the energy of the ground state zero-dimensional Laguerre-Gaussian mode (LG$_{00}$) such that it can be tuned through resonance with both $T_{PL}$ and $T_{RC}$. (d) Cavity PL intensity maps as the cavity mode is tuned through the trion resonances. Shown are the results at B $=0$ T (left panels) and B $=+8$ T (right panels) in both photon emission helicities. The laser is linearly polarized. At $B=0$ T, the spectra are essentially identical between both polarizations, while the near-unity spin polarization of the 2DEG at $B=+8$ T causes strong coupling to break down in $\sigma^-$ polarization. A modified coupled oscillator model incorporating the trion-polariton Stokes shift was used to fit the UPB and LPB (overlaid orange curves). }
\end{figure*} 

\begin{figure}
	\center
	\includegraphics[scale=1]{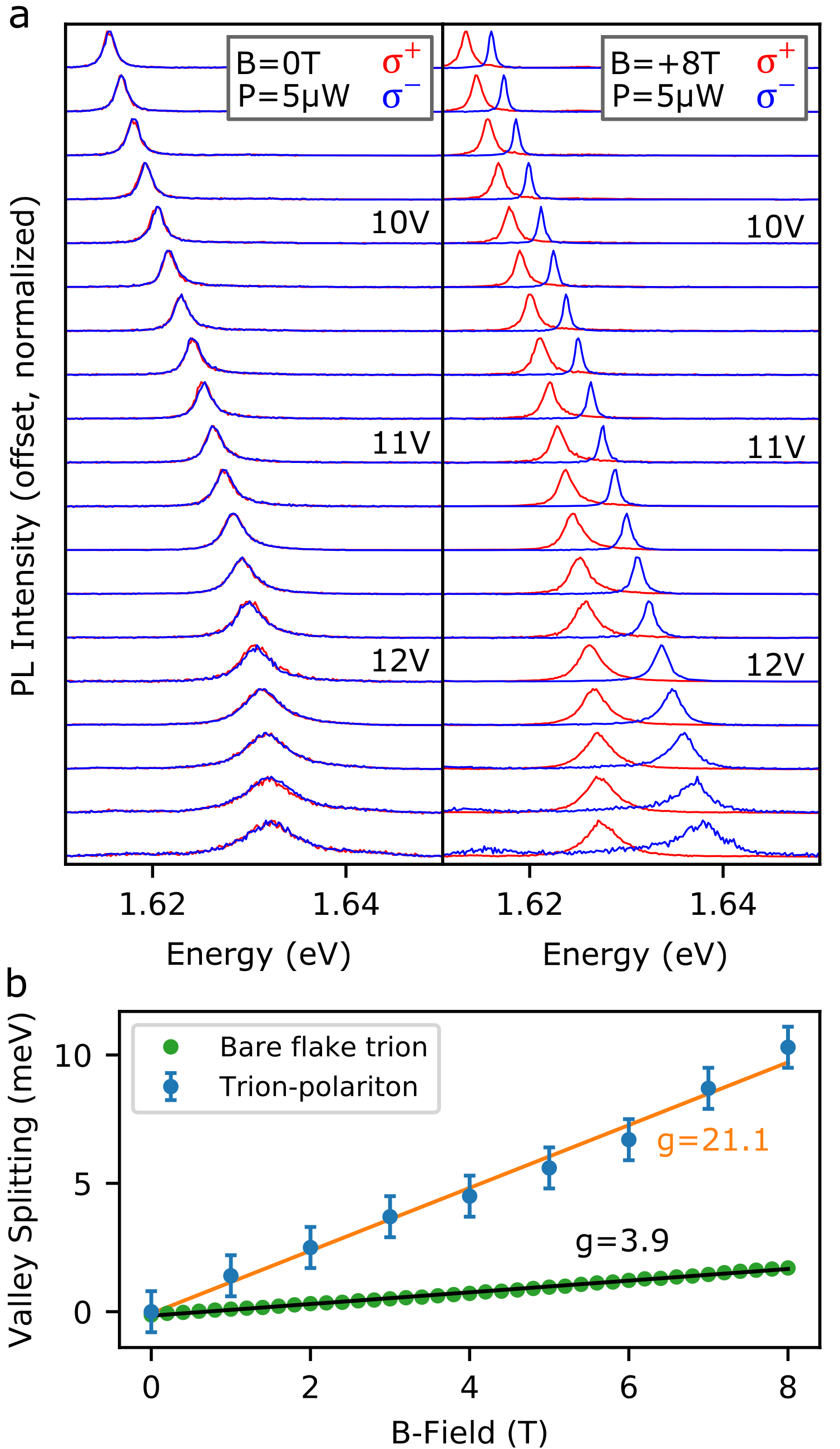}
	\caption{\label{fig:figure2} \textbf{Giant effective trion-polariton Zeeman splitting.} (a) Cavity PL spectra at increasing piezo voltages (decreasing cavity length) for B $=0$ T (left panel) and B $=+8$ T (right panel). A giant Zeeman splitting of the lower polariton branch (LPB) can be seen when the B-field is applied. (b) The maximum valley splitting of the trion-polariton LPB as a function of applied B-field strength. Here, we extract an effective maximum LPB Zeeman splitting at each 1 T B-field increment from our cavity fitting procedure (see Supplementary Note 2). For comparison the valley Zeeman splitting of the bare (uncoupled) trion is also shown. The g-factors of the trion-polariton and bare trion are $(21.1 \pm 0.9)$ and $(3.93 \pm 0.04)$, respectively. }
\end{figure}
 
\begin{figure*}
	\center
	\includegraphics[scale=1]{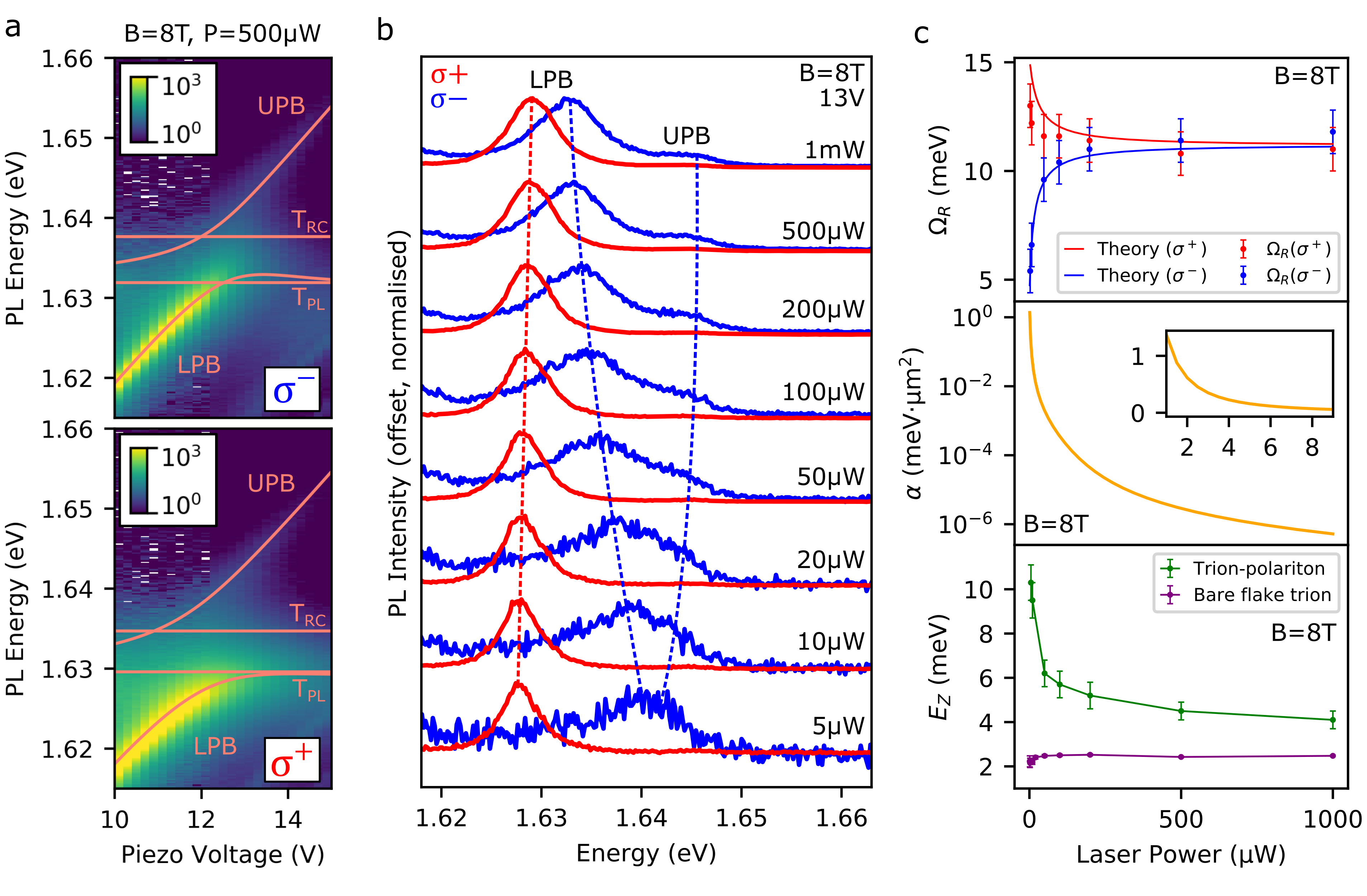}
	\caption{\label{fig:figure4} \textbf{Trion-polariton effective nonlinearity.} (a) Cavity PL colormaps in $\sigma^+$ and $\sigma^-$ emission at $B=+8$ T and a high laser power $P=500 \upmu$W. An anticrossing is seen in both polarizations despite the strong applied B-field. Polariton fitting curves incorporating the Stokes shift are overlaid. (b) Cavity PL spectra at fixed detuning close to trion-cavity resonance, at B $=+8$ T, taken at varying incident laser powers. As the power is decreased, the 2DEG spin polarization increases and the anticrossing in $\sigma^-$ is suppressed. This has the secondary effect of amplifying the effective Zeeman splitting between $\sigma^+$ and $\sigma^-$ lower polaritons. (c) (top panel) Rabi splittings in $\sigma^+$ and $\sigma^-$ at B $=+8$ T against laser power. Nonlinear breakdown of strong coupling in $\sigma^-$ is observed as the power is decreased. Solid curves are simulated results (see Supplementary Note 2). (middle panel) The calculated effective trion-polariton interaction strength $\alpha$ as a function of pump power at B $=+8$ T. Inset shows $\alpha$ at very low power. (lower panel) The maximum LPB Zeeman splitting, $E_Z$, at B $=+8$ T against laser power. The splitting increases drastically at the lowest powers when the 2DEG spin polarization is highest. For comparison the bare trion Zeeman splitting is shown, which is independent of laser power.}
\end{figure*}
 
Monolayer MoSe$_2$ presents a four-band massive Dirac system for studying spin and valley pseudospin dependent interactions between electrons, excitons, and photons \cite{Sidler2017,Back2017}. In the presence of an appreciable free carrier density, simple neutral exciton absorption evolves into two Fermi-polaron branches, repulsive and attractive \cite{Sidler2017,Back2017,Roch2019,Efimkin2017}. The monolayer then plays host to a Bose-Fermi mixture consisting of excitons dressed by electrons (or holes, for $p$-type doping). Strong coupling of these Fermi-polaron resonances to photonic microcavity modes has been demonstrated \cite{Sidler2017,BingTan2020}. Simplistically, the repulsive and attractive polarons correspond to a spin-triplet or spin-singlet interaction, respectively, between the two-dimensional electron gas (2DEG) and the constituent electron of the exciton \cite{Sidler2017,Back2017,Roch2019}. In MoSe$_2$, subject to strict spin-valley locking and chiral optical selection rules, this has the consequence of tying the 2DEG degree of spin polarization to the oscillator strengths of the polaron resonances in opposite photon helicities. The extreme example of this effect is when the 2DEG becomes fully spin polarized, leading to vanishing absorption of the attractive polaron in one photon helicity \cite{Back2017,Roch2019}.

It has recently been reported that when the Fermi level is significantly smaller than the trion binding energy, the attractive polaron may be adequately described as a three-body charged exciton, or trion \cite{Glazov2020,Imamoglu2021}. Although nominally the trion exists only in the strict single particle limit, in reality the transition between these two quasiparticle regimes is unclear, and likely depends heavily on the degree of exciton and carrier spatial localization over the monolayer, especially at low densities. This is particularly true in the case of nonequilibrium scenarios such as photoluminescence experiments, in which both species may coexist \cite{Imamoglu2021}.

Valley Zeeman splitting of these excitonic complexes has been reported under application of strong out-of-plane magnetic fields (B-fields) \cite{Back2017,Klein2009,Macneill2015}. However, translating the relatively large Zeeman splitting of a purely matter-bound excitation into a photonic mode splitting remains a fundamental challenge not only in opto-valleytronics \cite{langer2018lightwave}, but also in topological photonics. Indeed, many topological states of light have been implemented in recent years \cite{Ozawa2019}, including using TMD exciton-polaritons \cite{Khanikaev2020,Agarwal2020}. The ultimate goal of real topological protection against any type of disorder scattering and back-reflection requires time-reversal symmetry breaking \cite{wang2009observation,Lu2014}, with the size of the topological gap limited by the effective Zeeman splitting of the photonic modes. Large splittings are difficult to achieve at optical frequencies, and in the existing realizations either based on the use of magnetic proximity effects \cite{Bahari2017} or on the matter-based Zeeman splitting of exciton-polaritons \cite{Nalitov2015,Klembt2018}, the topological gap was $<1$~meV, too small to be clearly observable. 


In our work, by harnessing many-body interactions in a 2-dimensional Bose-Fermi mixture, we realise a giant effective trion-polariton Zeeman splitting, over 5 times larger than the bare (uncoupled) trion splitting, and more than double the polariton linewidths, a crucial step towards elimination of unwanted coupling between chiral modes \cite{Song2019a}. We moreover demonstrate giant effective non-linearity $\alpha\approx0.2\pm0.05$~meV$\cdot \upmu$m$^2$ for trion-polaritons under a magnetic field. This value is one order of magnitude larger than previously reported in TMDs \cite{Emmanuele2020,BingTan2020} and is based on an original mechanism involving free carrier valley relaxation and strong light-matter coupling. Robust photonic non-linearities, as in this work, are crucial for classical, quantum and topological photonics \cite{Lu2014,Ozawa2019}. 

We study a MoSe$_2$ monolayer on a 10 nm thick film of the ferromagnetic semiconductor europium sulphide (EuS) which coats a dielectric distributed Bragg reflector (DBR). Firstly, we characterize the MoSe$_2$ monolayer in the half-cavity, or bare flake, configuration, at temperature $T = 4.2$ K. Fig. 1a shows circular polarization resolved reflectance contrast (RC $=(R_0 - R)/R_0$, where $R$ and $R_0$ are the reflectance from the MoSe$_2$ and adjacent EuS substrate, respectively) spectra from the sample under linearly polarized broadband illumination at out-of-plane magnetic field strengths $B=-8,0,+8$ T. We observe, at $B = 0$ T, two clear absorption peaks attributed to the neutral exciton ($X_{RC}$) and trion ($T_{RC}$) at higher and lower energy, respectively. $T_{RC}$ displays a significant spectral weight, indicating an elevated doping level of the flake. These two resonances may be similarly described as Fermi-polarons, sharing the fundamental principle of a neutral exciton being either bound (attractive interaction, trion-like) or unbound (repulsive interaction) to itinerant carriers \cite{Efimkin2017,Sidler2017,Back2017,Roch2019}. The energy separation between these peaks allows us to estimate the free carrier density as $10^{12}$ cm$^{-2}$ (see Supplementary Note 1) \cite{Roch2019}. We attribute this relatively high carrier density to electron doping from the EuS film, which we expect to be highly charged owing to the deposition technique (see Methods) \cite{Keller2002,Grzeszczyk2020}. Measuring photoluminescence (PL) using a continuous wave laser at 1.946 eV, only a single peak is observed, attributed to the trion. The absence of neutral exciton PL is consistent with the high doping level in the flake. A significant Stokes shift of $\sim 6$ meV is observed between $T_{RC}$ and $T_{PL}$ (Fig. 1a).

When $B=\pm8$ T, $T_{RC}$ is only visible in one circular polarization (Fig. 1a). Owing to its spin-singlet or intervalley nature, the trion absorption strength of $\sigma^+$ ($\sigma^-$) light depends upon the itinerant carrier density in the $-K$ ($+K$) valley. Therefore, the electron Zeeman splitting is sufficiently large at this temperature to fully spin polarize the 2DEG (Fig. 1b) (see Supplementary Note 2) \cite{Back2017,Roch2019}. Achieving complete spin polarization of a 2DEG of such high density as here may point to itinerant ferromagnetism, in which transient domains of oppositely spin polarized electrons at $B=0$ T evolve into a spatially correlated spin polarized state when $B>0$ T \cite{roch2020first,lyons2020interplay}. We additionally note that while EuS is ferromagnetic, we see no evidence of magnetic proximity effects in the sample (see Supplementary Note 3).

For the next stage of the study, we incorporate the MoSe$_2$ / EuS structure into a tunable zero-dimensional microcavity (Fig. 1c), formed by introducing a downward facing top concave DBR into the optical path above the sample (as described in Ref. \cite{Dufferwiel2017}). By control of the mirror separation using piezo nanopositioners, we tune the ground state longitudinal cavity mode (Laguerre-Gaussian $LG_{00}$) through resonance with both $T_{PL}$ and $T_{RC}$, and perform cavity PL spectroscopy using a linearly polarized laser at power $5 \upmu$W. At $B=0$ T, we observe essentially identical PL spectra for both $\sigma^+$ and $\sigma^-$ detection polarizations. As the cavity length is tuned, the observation of an anticrossing indicates strong light-matter coupling and defines upper and lower trion-polariton branches (UPB and LPB) separated by a Rabi splitting $\Omega_R \sim 9$ meV. We note here that the trion Stokes shift is comparable with the Rabi splitting, and therefore must be taken into account in order to precisely fit the polariton PL energies by going beyond the most basic coupled oscillator model (see Supplementary Note 2). Indeed, while the anticrossing originates at the energy of $T_{RC}$, where cavity photons are most strongly absorbed, the polariton PL shows a finite Stokes shift causing both UPB and LPB emission to tend to the trion PL energy at vanishing photon fractions. Repeating the experiment at $B = +8$ T (Fig. 1d) reveals a larger anticrossing in $\sigma^+$, while the strong coupling regime breaks down in $\sigma^-$ ($\Omega_R$ is smaller than the polariton linewidths and unresolvable), consistent with the vanishing oscillator strength of $T_{RC}$ in $\sigma^-$ (Fig. 1a), and constituting observation of valley-specific strong light-matter coupling.

 

Fig. 2a shows $\sigma^+$ and $\sigma^-$ LPB PL versus piezo voltage at $B=0$ and 8 T, where a giant effective Zeeman splitting is observed, exceeding 10 meV as cavity length decreases. The LPB Zeeman splitting is amplified by valley-specific strong light-matter coupling, whereby the near-unity spin polarization of the 2DEG at B $=+8$ T suppresses the oscillator strength of the trion in $\sigma^-$ polarization, by transferring it to $\sigma^+$ polarization. Fig. 2b compares the trion PL g-factor measured on the bare flake ($g=3.9$) with that of the trion-polariton which is over 5 times larger ($g=21.1$). While the LPB Zeeman splitting increases at higher voltages, this comes at the cost of increased polariton linewidths and reduced intensity. However, we note that the LPB Zeeman splitting exceeds the bare trion splitting for all B-field strengths and all cavity lengths studied here. This result is in marked contrast to the expected scenario in which the polariton Zeeman splitting is reduced relative to that of bare trion by the corresponding Hopfield coefficient \cite{Lundt2019}.


Next, we show how the giant Zeeman splitting can be very effectively optically controlled. We fix $B = +8$ T and study the influence of incident laser power on the cavity PL. As can be seen in Fig. 3a, increased power reopens the anticrossing in $\sigma^-$ which previously collapsed upon application of the B-field (Fig. 1d). Fig. 3b shows trion-polariton PL spectra versus pumping power at fixed cavity length, where $\Omega_R$ grows in $\sigma^-$ and correspondingly decays in $\sigma^+$, suggesting that non-resonant pumping efficiently transfers electrons between spin-valley states. Here, qualitatively, electron-hole pairs are injected by the laser and bind to form excitons and trions on ultrafast timescales (sub-ps). The initial trion population will be highly valley polarized as the only free carriers available are from the spin polarized 2DEG, however, exciton and trion valley depolarization in MoSe$_2$ is extremely efficient (ps) owing to the Maialle-Silva-Sham (MSS) mechanism (confirmed here by transient ellipticity measurements, see Supplementary Note 4) \cite{Dufferwiel2017,Glazov2014}. Therefore, rapid intervalley scattering of trions followed by their radiative decay can result in a free electron remaining in the spin state anti-aligned to the external B-field. This means that each trion emission process results in partial transfer of electrons between spin-valley states. While trion valley relaxation occurs on ps timescales, the spin relaxation time for free electrons is $\sim1000$ times longer, of the order ns, as they are immune to the MSS mechanism and must undergo a large momentum transfer to scatter between spin-valley states. As such, trion intervalley scattering and subsequent photon emission can depolarize the 2DEG $\sim1000$ times faster than it can return to spin-polarized equilibrium. 
By embedding all of these processes into rate equations, we infer that laser power in the $\upmu$W range is enough to fully balance the 2DEG spin populations and associated trion-polariton Rabi splittings in opposite circular polarizations. Our simulations are shown in Fig. 3c (top panel) and are in excellent agreement with experimental data.

Lastly, we relate the computed exciton and trion densities to the energy shifts of both polariton branches, and deduce an effective interaction strength, which in this case is attractive for the LPB and repulsive for the UPB (Fig. 3c middle panel) (see Supplementary Note 2). The extracted value, $\alpha\approx0.2\pm0.05$~meV$\cdot \upmu$m$^2$ at $P=5~\mu$W, is one order of magnitude larger than previously reported for trion-polaritons because it is based on a completely different mechanism \cite{Emmanuele2020}. It is based neither on oscillator strength or the Coulomb interaction between carriers, but instead on linear spin relaxation processes. The increase in the interaction strength at the lowest laser powers is accompanied by a marked increase in the effective trion-polariton Zeeman splitting, confirming their shared origin in the 2DEG spin dynamics (Fig. 3c bottom panel).

Our experiments demonstrate the simultaneous manifestation of strong and weak coupling regimes between a photonic mode and a many-body correlated matter excitation consisting of an exciton dressed by electrons in an effective ferromagnetic phase, resulting in a giant Zeeman splitting between trion-polariton modes. We additionally show that laser illumination acts to depolarize the 2DEG via a process of trion valley pseudospin relaxation and subsequent radiative recombination. The resulting Rabi splitting transfer between the two polarization components induces energy renormalisation to which we associate large effective interactions. 
While in this work an EuS film was used to introduce additional free electrons into the flake, similar results should be observed in any MoSe$_2$ monolayer in which the itinerant carrier density can be raised arbitrarily to give the trion sufficient oscillator strength. Magnetic 2-dimensional materials may also be used to induce 2DEG spin polarization without the need for strong external B-fields \cite{lyons2020interplay}. Moreover, we note that extremely high laser powers, often pulsed and quasi-resonant, are typically needed to enter regimes of polariton non-linearity, while here the strongest effective interactions occur under low power non-resonant continuous-wave laser excitation. Our work therefore highlights doped MoSe$_2$ as a flexible system in which to realize and apply ultrastrong low-threshold non-linearities, for instance towards TMD-based all-optical logic gates \cite{Amo2010}, or to explore nonlinear topological photonics \cite{smirnova2020nonlinear}.



\section{Methods}

\subsection{Low temperature magneto-optical spectroscopy}

Magneto-optical spectroscopy at 4.2 K was performed by mounting the sample in a liquid helium bath cryostat with a superconducting magnet and free space optical access. Reflectance contrast measurements were performed by directing broadband white light in either $\sigma^+$ or $\sigma^-$ circular polarization onto the sample and measuring the reflected signal on the MoSe$_2$ monolayer ($R$) and adjacent bare EuS film ($R_0$), and calculating the RC $= \Delta R/R$. Photoluminescence spectroscopy was performed by directing a linearly polarized continuous wave laser at 1.946 eV onto the sample and detecting the emission in either $\sigma^+$ or $\sigma^-$ circular polarization. For both RC and PL the signal was directed through a single mode fiber to a 0.75 m spectrometer and onto a nitrogen-cooled high sensitivity charge-coupled device.

\subsection{Europium sulphide deposition}

A 10 nm thick film of europium sulfide (EuS) was deposited onto a dielectric DBR (top layer SiO$_2$) by electron-beam evaporation. By maintaining a low substrate temperature of 16 $^\circ$C during the deposition, we ensure that the resulting EuS film will be sulfur deficient, owing to the much lower vapor pressure of S relative to Eu, causing S atoms to re-evaporate from the substrate during growth. The resulting sulfur vacancies act as electron donors causing the non-stoichiometric EuS film to act as a heavily-doped ferromagnetic semiconductor \cite{Keller2002}. The MoSe$_2$ monolayer therefore becomes highly charged when it is stamped on top of the EuS substrate \cite{Grzeszczyk2020}. 

\subsection{Sample fabrication}

A MoSe$_2$ bulk crystal supplied by HQ Graphene was exfoliated with tape onto a polydimethylsiloxane (PDMS) sheet, and a suitable monolayer identified by optical microscopy. This monolayer was then stamped onto the DBR / EuS substrate using a conventional viscoelastic dry transfer method.

\section{Acknowledgements}

TPL acknowledges financial support from the EPSRC Doctoral Prize Fellowship scheme. TPL, DJG, JP, YO and AIT acknowledge support from the Royal Society International Exchange Grant IEC\textbackslash R3\textbackslash 170088. TPL, DJG, AG, CLo, LK, IA, MB and AIT acknowledge EPSRC Centre-to-Centre grant EP/S030751/1. TPL, DJG and AIT additionally acknowledge financial support of the European Graphene Flagship Project under grant agreement 881603 and EPSRC grants EP/V006975/1 and EP/P026850/1. CLe, DS and GM acknowledge the support of the projects EU ``TOPOLIGHT" (964770), ``QUANTOPOL" (846353), of the ANR Labex GaNEXT (ANR-11-LABX-0014), and of the ANR program ``Investissements d'Avenir" through the IDEX-ISITE initiative 16-IDEX-0001 (CAP 20-25). The Dortmund team acknowledges financial support by the Deutsche Forschungsgemeinschaft through the International Collaborative Research Centre 160 (Project No. C2) and UAR professorship: Mercur Foundation (grant Pe-2019-0022). The authors thank D. N. Krizhanovskii for useful discussions.

\section{Author contributions}

TPL, DJG and JP performed low temperature magneto-optical spectroscopy. TPL, DJG, CLe, DDS, GM and AIT analyzed and discussed the bare flake and cavity spectroscopy data. CLe, DDS and GM developed the cavity fitting model and rate equations. LK and IAA collected and analyzed time-resolved data. JP and PM deposited the EuS films onto DBR substrates. TPL, DJG, JP and PM performed SQUID magnetometry. CLo identified and transferred MoSe$_2$ flakes onto EuS films. AG carried out electron density calculations. MB, YO, GM and AIT managed various aspects of the project. AIT supervised the project. TPL wrote the manuscript with contributions from all co-authors.

\end{document}


\title{Supplementary Information for: Giant effective Zeeman splitting in a monolayer semiconductor realized by spin-selective strong light-matter coupling}

\author{T. P. Lyons}
\email{t.lyons@sheffield.ac.uk}
\author{D. J. Gillard}
\affiliation{Department of Physics and Astronomy, The University of Sheffield, Sheffield S3 7RH, UK}
\author{C. Leblanc}
\affiliation{Institut Pascal, PHOTON-N2, CNRS, Universit\'e Clermont Auvergne, F63000 Clermont-Ferrand, France}
\author{J. Puebla}
\affiliation{Center for Emergent Matter Science, RIKEN, Wako 351-0198, Japan}
\author{D. D. Solnyshkov}
\affiliation{Institut Pascal, PHOTON-N2, CNRS, Universit\'e Clermont Auvergne, F63000 Clermont-Ferrand, France}
\affiliation{Institut Universitaire de France (IUF), F-75231 Paris, France}
\author{L. Klompmaker}
\affiliation{Experimentelle Physik 2, Technische Universit\"at Dortmund, 44221 Dortmund, Germany}
\author{I. A. Akimov}
\affiliation{Experimentelle Physik 2, Technische Universit\"at Dortmund, 44221 Dortmund, Germany}
\affiliation{Ioffe Institute, Russian Academy of Sciences, 194021 St. Petersburg, Russia}
\author{C. Louca}
\affiliation{Department of Physics and Astronomy, The University of Sheffield, Sheffield S3 7RH, UK}
\author{P. Muduli}
\affiliation{Institute for Solid State Physics, University of Tokyo, Kashiwa 277-8581, Japan}
\affiliation{Department of Physics, Indian Institute of Technology Madras, Chennai 600036, India}
\author{A. Genco}
\affiliation{Department of Physics and Astronomy, The University of Sheffield, Sheffield S3 7RH, UK}
\author{M. Bayer}
\affiliation{Experimentelle Physik 2, Technische Universit\"at Dortmund, 44221 Dortmund, Germany}
\affiliation{Ioffe Institute, Russian Academy of Sciences, 194021 St. Petersburg, Russia}
\author{Y. Otani}
\affiliation{Center for Emergent Matter Science, RIKEN, Wako 351-0198, Japan}
\affiliation{Institute for Solid State Physics, University of Tokyo, Kashiwa 277-8581, Japan}
\author{G. Malpuech}
\affiliation{Institut Pascal, PHOTON-N2, CNRS, Universit\'e Clermont Auvergne, F63000 Clermont-Ferrand, France}
\author{A. I. Tartakovskii}
\email{a.tartakovskii@sheffield.ac.uk}
\affiliation{Department of Physics and Astronomy, The University of Sheffield, Sheffield S3 7RH, UK}
\maketitle

\clearpage

\subsection{\textbf{Supplementary Note 1: Electron density calculations}}

We estimate the free electron density in our MoSe$_2$ monolayer by following the treatment described by Roch, \textit{et al.} for MoS$_2$ \cite{Roch2019}, itself motivated by earlier works by Suris, \textit{et al.} \cite{Suris2001}, in which exciton absorption in the presence of a 2DEG was considered. In this treatment, the absorption of the system may be described by two exciton-electron scattering mechanisms, that of a singlet or triplet interaction between the 2DEG and the photogenerated exciton. Two scattering matrix elements, $T_s$ and $T_t$ respectively, can therefore be written as \cite{Roch2019,Suris2001}:

\begin{eqnarray}
T_s(\hbar \omega + E) &=& \frac{2 \pi \hbar ^2}{\mu_T} \frac{1}{\ln \Big(\frac{-E_b (X^-)}{(\hbar \omega + E) - E(X^0) + i\gamma} \Big)} \\
T_t(\hbar \omega + E) &=& \frac{2 \pi \hbar ^2}{\mu_T} \frac{1}{\ln \Big( \frac{-E_b (X^0)}{(\hbar \omega + E) - E(X^0) + i \gamma} \frac{E_b (X^0)}{E_b (X^-)} \Big)}
\end{eqnarray}

\noindent where $E_b (X^0)$ and $E_b (X^-)$ are the exciton and trion binding energies, $E(X^0)$ is the exciton energy, $\gamma$ is the exciton broadening, and $\mu_T$ is the reduced trion mass \cite{Roch2019}:

\begin{eqnarray}
\frac{1}{\mu_T} = \frac{1}{m_{CB}} + \frac{1}{m_{CB}+m_{VB}}
\end{eqnarray}

\noindent where $m_{CB}$ and $m_{VB}$ are the effective conduction and valence band masses. The scattering matrices combine as $T = \frac{1}{2}T_s + \frac{3}{2}T_t$, from which we can calculate the self energy $\Xi$ for the case of $E_F < E_b(X^-)$ \cite{Roch2019,Suris2001}:

\begin{eqnarray}
\Xi(\hbar \omega, E_F) = \int_{0}^{\infty} g_{2D} f_{FD}(E, E_F) T(\hbar \omega + E) \mathrm{d}E
\end{eqnarray}

\noindent where $g_{2D}$ is the 2D density of states (without spin degeneracy) $m_{CB} / 2(\pi \hbar^2)$ and $f_{FD}$ is the Fermi-Dirac distribution:

\begin{eqnarray}
f_{FD}(E, E_F) = \frac{1}{\exp(\frac{E-E_F}{k_B T})+1}
\end{eqnarray} 

\clearpage 

From the self energy we approximate the optical susceptibility \cite{Roch2019,Suris2001}:

\begin{eqnarray}
\chi(\hbar \omega, E_F) \approx \frac{-A}{\hbar \omega + i \gamma - E(X^0) - \Xi}
\end{eqnarray} 

\noindent where A is a term accounting for the intervalley optical dipole moment and the exciton wavefunction at $r=0$, being roughly equivalent to 0.1 in our case \cite{Roch2019}. 

To simulate the absorption of the MoSe$_2$ we plot the imaginary part of the susceptibility in Fig. 1a as a funciton of Fermi level, using $E_b (X^0) = 550$ meV \cite{Ugeda2014}, $E_b (X^-) = 25$ meV, $E(X^0) = 1.668$ eV, $\gamma = 3$ meV, $T=4$ K, $m_{CB} = 0.49$ and $m_{VB} = 0.52$ \cite{Wang2015a}. It can be seen that as $E_F$ increases, the energy separation between the upper and lower polaron branches increases (Fig. 1b). From the Main Text Fig. 1a, we extract the energy separation at $B=0$ T between the repulsive and attractive polaron branches as 32.5 meV, and compare this value to Fig. 1b in order to obtain the Fermi level in our MoSe$_2$ monolayer, which we find to be $E_F = (5 \pm 3)$ meV. The large error in $E_F$ arises from the large uncertainty in the trion binding energy in literature, which varies from 21 meV \cite{Berkelbach2013} to 30 meV \cite{Ross2013}. Here we take an intermediate value $E_b(X^-) = 25$ meV as used by Sidler, \textit{et al.} \cite{Sidler2017}. Finally, we estimate the free electron density by the 2D density of states with spin degeneracy, $n_e = (m_{CB}m_e E_F) / (\pi \hbar^2)$ and obtain a value $(1.0 \pm 0.6) \times 10^{12}$ cm$^{-2}$.

\begin{figure*}[h]
\begin{center}
\includegraphics[scale=1]{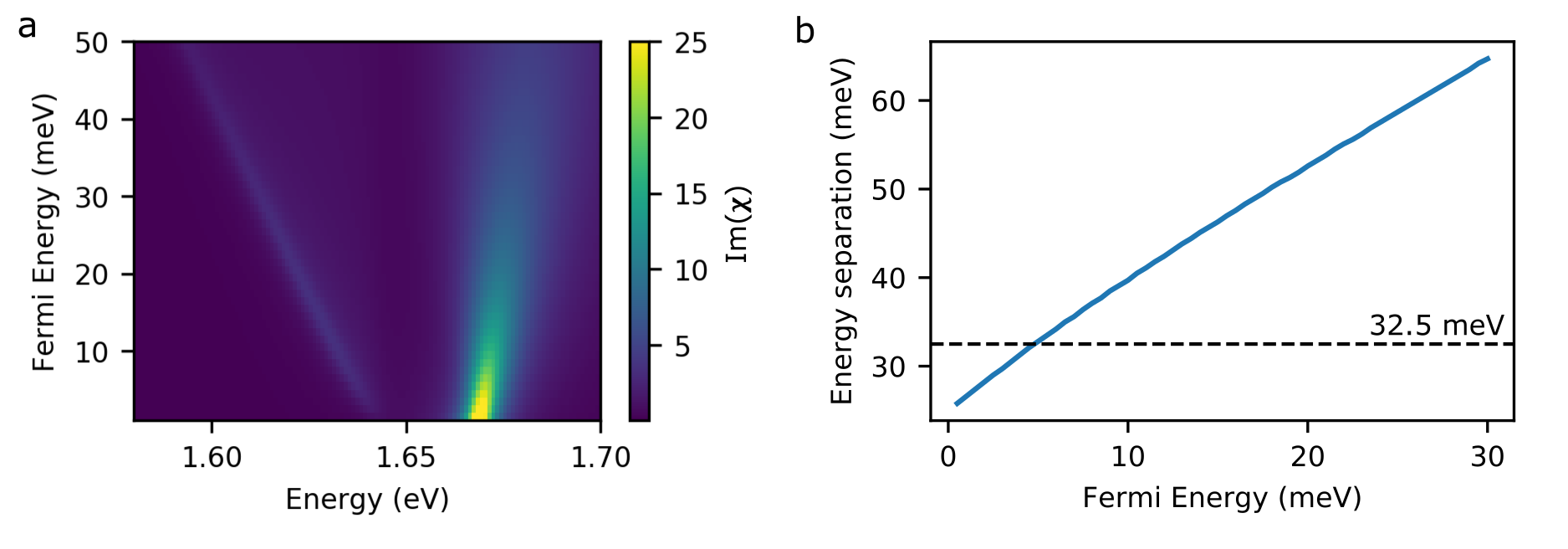}
\end{center}
\caption{(a) Calculated imaginary part of the MoSe$_2$ optical susceptibility at $B=0$ T as a function of increasing Fermi energy.  (b) Energy separation between the exciton and trion (repulsive and attractive polaron branches) as a function of Fermi energy. In our sample the energy separation is 32.5 meV, corresponding to $E_F \sim 5$ meV.}
\end{figure*}

\clearpage

\subsection{\textbf{Supplementary Note 2: Theory of the spin dynamics}}

\section{Photoluminescence from polariton modes}

We theoretically describe each circularly polarised trion-polariton mode using a standard two-coupled oscillators model. Since the broadening is significant, we include mode broadening by using an homogenous imaginary part for the bare photon and trion energies \cite{SAVONA1995,CavityPolariton2003}. The trion energies are determined by reflection experiments performed on the bare flake. The open cavity mode energy reads as $E_c=\alpha(V+V_0)$ where $\alpha$ the slope of the photonic mode energy versus the applied voltage $V$ controlling the optical cavity length.
The upper and lower polariton branch energies read as :

\begin{eqnarray}
 {E_{U,LPB}^{\sigma_{\pm}}} = \frac{1}{2}\operatorname{Re} \left( {E_{Ta}^{\sigma_{\pm}} + {E_c} \pm \sqrt {\left( {E_{Ta}^{\sigma_{\pm}} - {E_c} + i\left( {{\Gamma _{Ta}} - {\Gamma _{c}} } \right)} \right)^2+4({\Omega^{\sigma_{\pm}})^2}}} \right)\label{polEn}
\end{eqnarray}

\noindent where $E_{Ta}^{\sigma_{\pm}}$ is the energy of the trion in absorption for a given spin component,  $\frac{\Omega^{\sigma_{\pm}}}{2}$ are the Rabi splitting values to be determined, $\Gamma_{c}$=0.6 meV and $\Gamma_{Ta}$=6 meV are the measured linewidth of the two modes, which we keep constant.

The trion fractions of the polaritons  (square of the Hopfield coefficients)  read as:

\begin{eqnarray}
  \abs{X_L^{\sigma_{\pm}}}^2 &=&\frac{{4(\Omega^{\sigma_{\pm}})^2 }}{{4{\Omega ^2} + \left( {{E_{UPB}^{\sigma_{\pm}}} - E_{Ta}^{\sigma_{\pm}}} \right)} }  \\ \abs{X_U^{\sigma_{\pm}}}^2&=&1-\abs{X_L^{\sigma_{\pm}}}^2
\end{eqnarray}

\begin{figure*}[tb]
\begin{center}
\includegraphics[width=\linewidth]{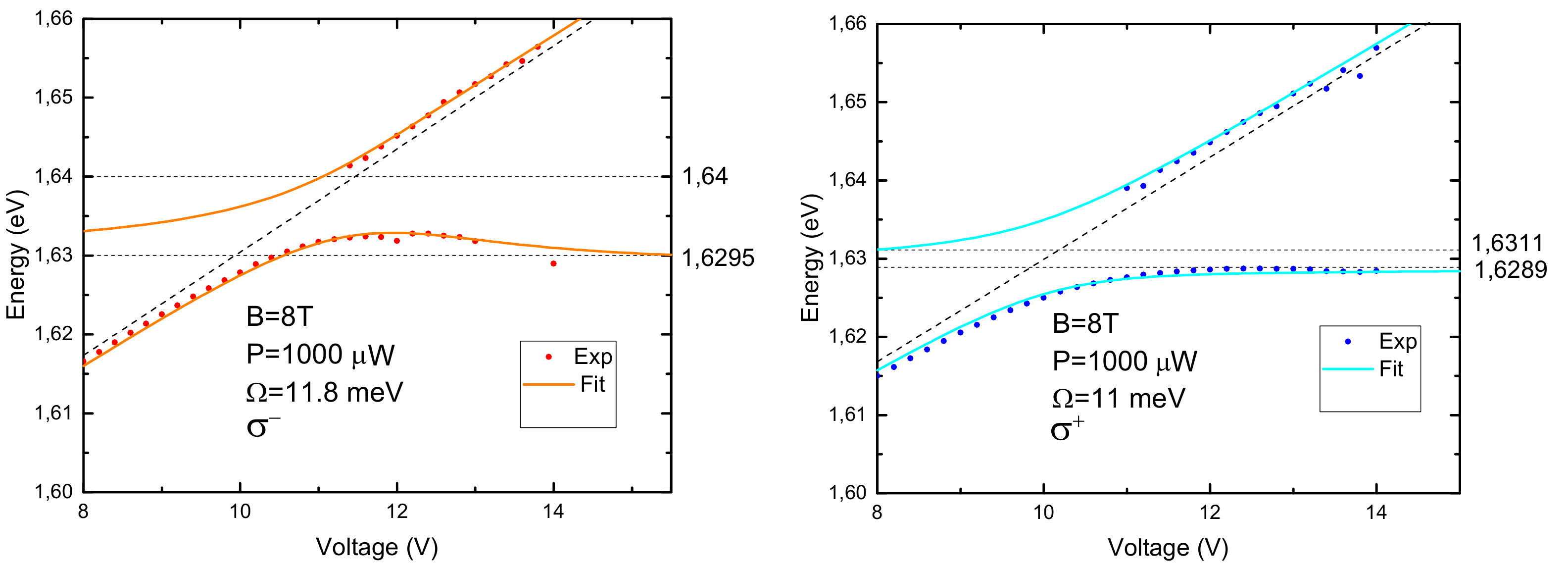}
\caption{Fit of the lower and upper polariton branches in two polarizations: dots -- experiment, line -- theory.\label{fig3}}
\end{center}
\end{figure*}

These energies are the one which could be measured in a transmission or absorption experiment. As for bare excitons in any media, polaritons demonstrate a finite Stokes shift. In the presence of finite random disorder, the Stokes shift is the difference between the energy of a resonance in absorption which corresponds to the maximum density of states, and in photoluminescence (PL) which corresponds to the lowest energy state of the inhomogeneous distribution of energy resonances. As shown in Fig.~1a of the main text, we have performed both reflection and PL measurements on the bare flakes which allows us to precisely determine the bare trion Stokes shift value, which is of the order of 6~meV. It reads

\begin{equation}
    \Delta_S^{\sigma_{\pm}} = E_{Ta}^{\sigma_{\pm}} - E_{Te}^{\sigma_{\pm}}
\end{equation}

\noindent where $E_{Te}^{\sigma_{\pm}}$ is the trion energy in PL.

For polaritons, the Stokes shift is reduced with respect the case of the bare trion. The polariton Stokes shift should tend to the trion Stokes shift when the trion fraction of the polariton tends to 1. On the other hand, the Stokes shift should tend to 0 
when the polariton becomes strongly photonic. We therefore choose to introduce phenomenologically a polariton Stokes shift value given by:

\begin{eqnarray}
  {E_{U,L}^{\sigma_{\pm}}}& = &{E_{U,LPB}^{\sigma_{\pm}}} - \Delta_S^{\sigma_{\pm}}\ |{X_{U,L}^{\sigma_{\pm}}}|^2  
\end{eqnarray}

Using the above mentioned formula allows us to fit the voltage-dependence of the trion-polariton PL energies for different pumping powers and magnetic field strengths, with the Rabi splitting in each polarisation being the only fitting parameters. The results are shown in Fig.~1d of the main text and Fig.~2 of the supplementary. The agreement between the experiment and the phenomenological theory which we use is extremely satisfactory. The extracted dependence of the Rabi splitting values versus pumping for both spin components are shown on Fig.~3c of the main text.



\section{Modelling of free carrier depolarisation}
In this section we present the system of rate equations we use in order to compute the free carrier spin polarisation versus pumping power. This model can be used, with different parameters, to describe the relaxation dynamics both in a bare flake and when the flake is embedded in the cavity. In this last case, the computed polarisation degree of carriers (free and those bound to excitons to form trions) allows to directly deduce the dependence of Rabi splitting  values versus pumping, and to compare them with the experiment. 
The scheme of the processes taking place in the system is shown in Suppl. Fig.~3. When pumping starts, the resident carriers are assumed to be all spin-polarized up, because of the applied magnetic field. The pump creates excitons (exciton-polaritons), which bind with the free carriers to form trions (trion-polaritons) with correlated spin. Trions can depolarize a lot faster ($\sim$ ps) than free carriers ($\sim$ ns) because of the L-T spin-orbit coupling (Maialle-Sham mechanism\cite{Maialle1993,Glazov2014}). When these depolarized trions emit light via the exciton recombination and emission of the photon out of the cavity, the remaining free carriers have a reduced polarization.  This reduced polarisation depends on the ratio between the trion (trion-polariton) decay time and the trion depolarisation time. The time needed for an electron spin to reorient parallel to the magnetic field is comparable with its depolarisation time ($\sim$ ns). Therefore, even with a small pumping (a few excitons per picosecond), the small number of trions formed by this process can efficiently depolarize a large fraction of free carriers. 

\begin{figure}[b]
\begin{center}
\includegraphics[width=0.7\linewidth]{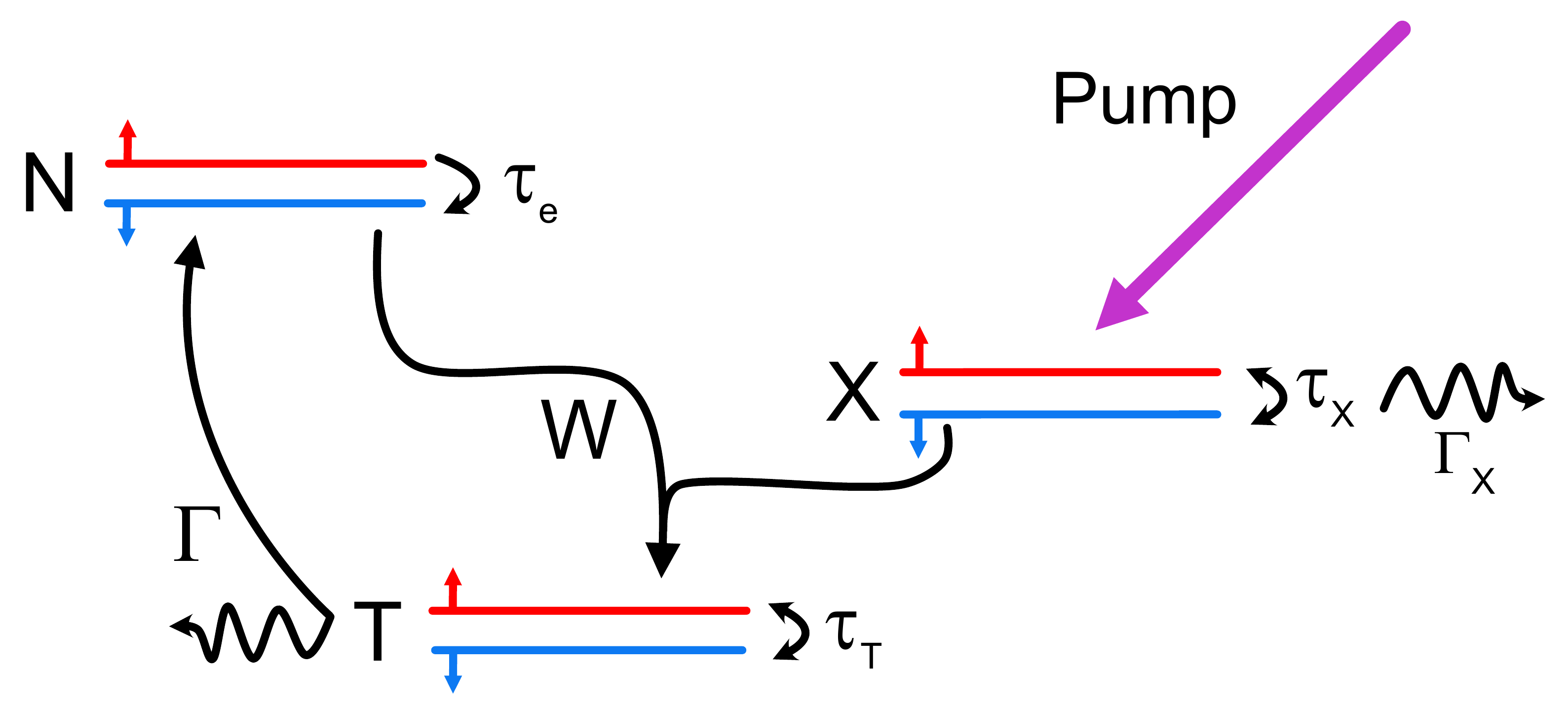}
\caption{Scheme of the population dynamics. Arrows indicate the pumping (violet), scattering and decay rates (black), spin relaxation rates (small black). \label{fig1}}
\end{center}
\end{figure}

The result is that, under optical pumping, we observe strong coupling in a cavity for both polarizations, because there are free carriers of both spins available in the system. The key point is that the effect of the injected exciton density on the Rabi splitting is amplified by the ratio $\tau^s_e/\tau^s_t\sim 10^3$.

Mathematically, the processes described can be modelled by rate equations involving free carriers $N$, excitons (exciton-polaritons) $X$, and trions (trion-polaritons) $T$:

\begin{eqnarray}
\frac{{d{N_ \pm }}}{{dt}}& =&  - W{N_ \pm }{X_ \pm } + \Gamma {T_ \pm } \pm \frac{{{N_ - }}}{{{\tau _e^s}}} \mp \frac{{{N_ + }}}{{{\tau _e^s}}}{e^{ - \frac{\Delta }{{{k_b}T_e}}}}\\
\frac{{d{X_ \pm }}}{{dt}}& =&  - W{N_ \pm }{X_ \pm } + {P_ \pm } - \frac{{{X_ \pm } - {X_ \mp }}}{{{\tau _X^s}}} - \frac{X_ \pm }{\tau _X}\\
\frac{{d{T_ \pm }}}{{dt}} &= &W{N_ \pm }{X_ \pm } - \Gamma {T_ \pm } - \frac{{{T_ \pm } - {T_ \mp }}}{{{\tau _T^s}}}
\end{eqnarray}


The $\pm$ indices of $N$, $X$, and $T$ correspond to the different spins of the free carriers, excitons (exciton-polaritons), and trions (trion-polaritons), respectively.$\Delta$ is the electron Zeeman splitting, $T_e$ is the temperature. In all cases, we have considered linearly-polarized pumping: $P_+=P_-$.  The initial number of spin-polarized free carriers is $N_0=1000$ (estimated from the density $10^{12}$~cm$^{-2}$).
The table below gives the parameters we used to describe the bare flake under 0 and 8~T and the cavity system. The trion-polariton decay rate toward a photon and a free electron is $\Gamma=1/\tau_T$, where $\tau_T$ is the trion-polariton's lifetime.


\begin{center}
\begin{tabular}{|c|| c | c |c | }
 \hline
  \multicolumn{4}{|c|}{ Table of parameters} \\
 \hline			
 Parameter & Bare flake 0T  &Bare flake 8T & Cavity 8T \\ 
  & Measured experimentally & Fit parameters & Fit parameters \\ \hline 
   $\tau_T$   &    & 0.2~ps & 2~ps \\ \hline
   $\tau_T^s$ & 1~ps & 2.7~ps & 10~ps \\ \hline
   $\tau_X$   &    & 0.3~ps & 2~ps \\ \hline
   $\tau_X^s$ & 0.2~ps   & 0.7~ps & 5~ps \\ \hline
   $\tau_e^s$ & 3~ns   & 3~ns   & 3~ns \\ \hline
   $\Delta$ & 0 meV  & 3.4~meV & 3.4~meV \\ \hline
   $W$ &   & 3x10$^{-3}$~ps$^{-1}$ & 3x10$^{-3}$~ps$^{-1}$ \\ 
 \hline  
 \end{tabular}
 \end{center}

The spin lifetimes in the bare flake at 0T have been measured by time resolved spectroscopy, as described in Supplementary Note 4. The parameters used to describe the bare flake at 8~T have been used to compute the power dependence of the DOCP. The exciton and trion lifetimes correspond to their radiative lifetime, which is proportional to their oscillator strength. This quantity is directly measured from the Rabi splitting value when the flake is embedded in the cavity. Formally, this quantity should depend on the free carrier spin polarisation, but this dependency is neglected. 
The spin relaxation time of excitons and trions are expected to be longer at 8T than at 0~T, due to the protection offered by the Zeeman splitting. The trion DOCP at low power (around 0.85-0.9) is strongly fixing the ratio between the trion spin lifetime and its radiative lifetime. Suppl. Fig.~4 shows the trion DOCP versus pumping, demonstrating a good agreement between theory and experiment, where increased pumping provokes a depolarisation of the free carrier gas and a decrease of the DOCP.

In the cavity, the lifetime of the strongly coupled polariton modes is not anymore given by the radiative lifetime of excitons and trions, which now sets the Rabi splitting values. The trion-polariton and exciton-polariton lifetimes are composed by the decay of the photonic part of the polariton (cavity photon lifetime ~1~ps) and the non-radiative decay of excitons/trions. In line with our previous studies \cite{dufferwiel2017valley}, we consider a polariton being 50 \% photon with a decay rate dominated by the cavity photon lifetime. We also find spin relaxation times substantially larger than in the bare flake. The explanation suggested in \cite{dufferwiel2017valley} is that exciton/trion-polaritons in 0D cavities are constituted by low wavevector excitons with small L-T splitting, which slows down their spin relaxation via to the Maialle-Sham mechanism with respect to a spatially invariant 2D system. In the same way, the elastic disorder scattering is strongly suppressed with respect to the bare trion case, which reduces the disorder contribution to the decay and makes the polariton decay rate dominated by the photonic part of the polariton.

With these parameters we compute the steady state populations of all species versus pumping power. This allows to determine the oscillator strength of the bare trion resonance for each spin, which is proportional to the density of carriers with the same spin \cite{emmanuele2020highly}. The Rabi splitting is proportional to the square root of the oscillator strength \cite{CavityPolariton2003}. For a given polarisation, it is therefore proportional to the square root of the number of carriers of the corresponding spin (both free and bound in trions). Fig.~3c of the main text shows the change of the Rabi splitting of each polarisation versus pumping computed with the above mentioned model compared with the experimentally extracted data. The agreement is very satisfactory. Finally, we compute the effective trion-polariton interaction strength. By definition, the interaction constant is given by $\alpha={\partial E_{LPB}^+}/{\partial n^+}$, where $E_{LPB}^+$ is the lower polariton branch energy for the $\sigma^+$ polarization calculated with equation \eqref{polEn} and $n^+$ is the particle density corresponding to the pumping value $P$. This correspondence is determined by the free electron density, discussed in the Supplementary Note 1, and by the fitting parameters, summarized in the Table above. Fig.~3c (middle panel) of the main text shows the evolution of the effective interaction $\alpha$ versus the pumping power for a given magnetic field $B=+8T$. The high value of the interaction constant is qualitatively explained by the high ratio of the spin relaxation times, allowing a small density of excitons to depolarize the free electrons thus strongly affecting the polariton energy via the Rabi splitting.
\newline

\begin{figure*}[h]
	\begin{center}
		\includegraphics[scale=0.5]{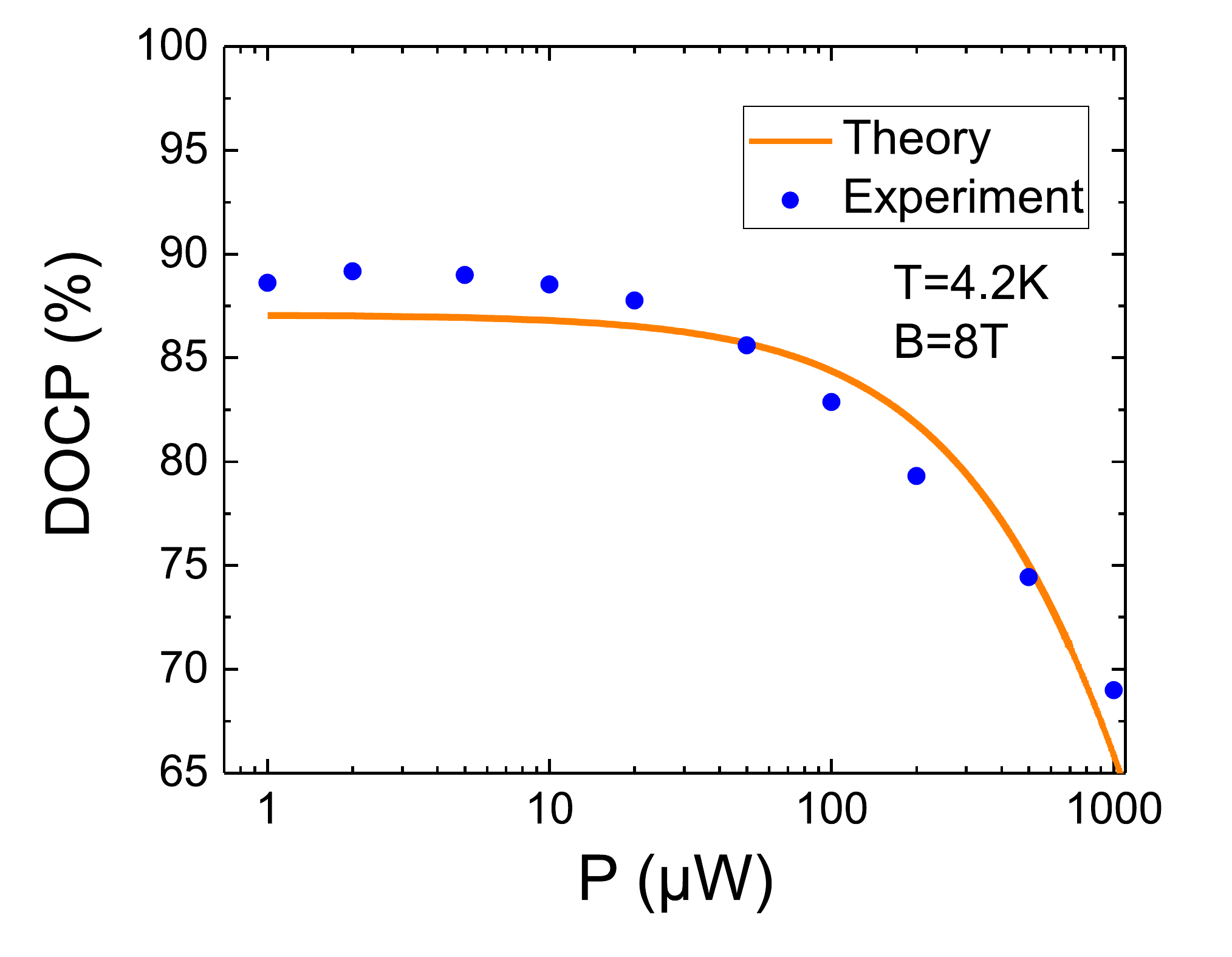}
	\end{center}
	\caption{Power dependence of the bare flake trion DOCP in photoluminescence at $B=8$ T. Overlaid is the simulated response.}
\end{figure*}

\clearpage

\subsection{\textbf{Supplementary Note 3: Enhanced valley splitting of MoSe$_2$ on EuS substrates}}

\begin{figure*}[h]
	\begin{center}
		\includegraphics[scale=1]{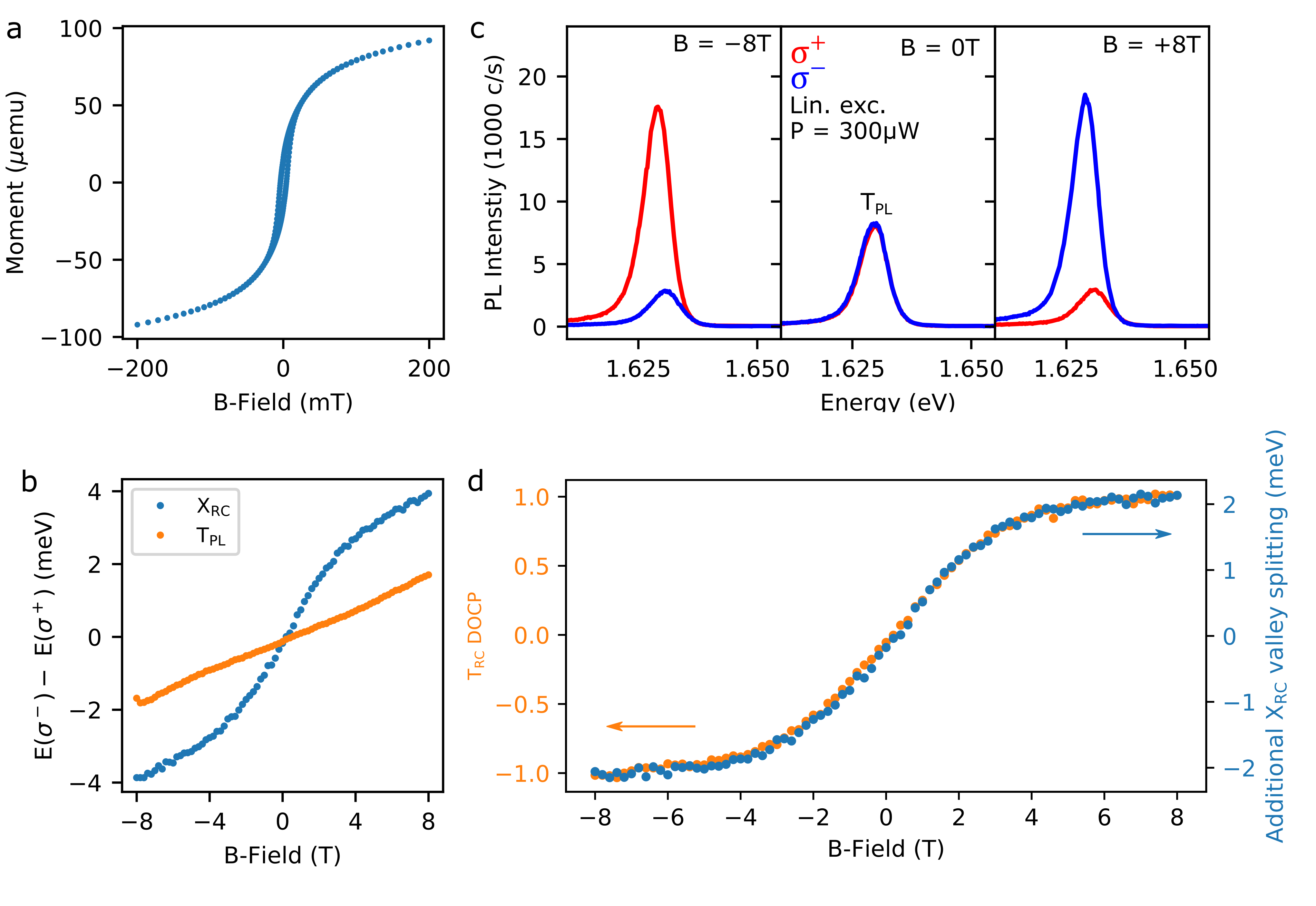}
	\end{center}
	\caption{(a) Vibrating-sample magnetometer (VSM) data from the EuS film used in this work, showing clear signatures of ferromagnetism. The B-field is applied in the sample plane and the sample temperature is 4 K. (b) The valley splitting (here defined as $E(\sigma^-) - E(\sigma^+)$) of $X_{RC}$ and $T_{PL}$, showing that while $X_{RC}$ displays an enhanced and non-linear valley splitting, $T_{PL}$ does not. (c) Trion PL spectra from the bare flake at B$=-8, 0, +8$ T, under non-resonant laser excitation. (d) Comparison between the enhancement to the $X_{RC}$ valley splitting (calculated as the measured $X_{RC}$ valley splitting minus a linear shift of 0.23 meV / T arising from the valley Zeeman effect caused by the externally applied B-field) and the DOCP of $T_{RC}$. The exact match between the two B-field responses supports the conclusion that both effects arise from the same origin, that of 2DEG redistribution between valley states.}
	\label{EuS_fig}
\end{figure*}

The sample used in this work consists of a MoSe$_2$ monolayer in direct contact with a 10 nm thick europium sulfide (EuS) thin film. We confirm the ferromagnetism of this film by vibrating-sample magnetometry with an in-plane B-field, where clear magnetization saturation and hysteresis is seen (Fig.~\ref{EuS_fig}a). In recent years, monolayer TMDs on EuS substrates have been shown to exhibit an enhanced valley Zeeman splitting in the fundamental exciton absorption peak, attributed to the interfacial magnetic exchange field \cite{Norden2019,Zhao2017,Scharf2017}. In our sample, an enhanced and nonlinear valley splitting of the neutral exciton $X_{RC}$ (repulsive polaron) is also observed, however, the enhancement is not seen in the valley splitting of the trion PL peak $T_{PL}$, which shifts with applied field at a rate of 0.23 meV / T, consistent with the conventional valley Zeeman effect in MoSe$_2$ (Fig.~\ref{EuS_fig}b) \cite{Macneill2015}. As the PL valley splitting is a better reflection of the energy shifts of the band edges under a B-field than the RC valley splitting, we rule out an interfacial exchange field induced valley splitting of $+K$ and $-K$ band edges. Instead, we interpret the enhanced valley splitting seen in $X_{RC}$ as being due to state filling as the raised free carrier population in the flake redistributes between conduction band valleys, thereby enlarging the photon absorption energy. This is additionally supported when we consider the enhancement to the valley splitting of $X_{RC}$, by subtracting a linear shift of 0.23 meV / T from the measured splitting. This is plotted in Fig.~\ref{EuS_fig}d overlaid with the DOCP of $T_{RC}$, where it can be seen that both curves follow exactly the same response to the applied external B-field. This strongly suggests that the enhancement to the valley splitting of $X_{RC}$ and the DOCP of $T_{RC}$ both arise from the redistribution of free electrons between the conduction band valleys as the valley degeneracy is lifted by the applied B-field.

\clearpage

\subsection{\textbf{Supplementary Note 4: Time resolved spectroscopy of monolayer MoSe$_2$ on EuS substrates}}

\begin{figure*}[h]
	\begin{center}
		\includegraphics[width=1.0\textwidth]{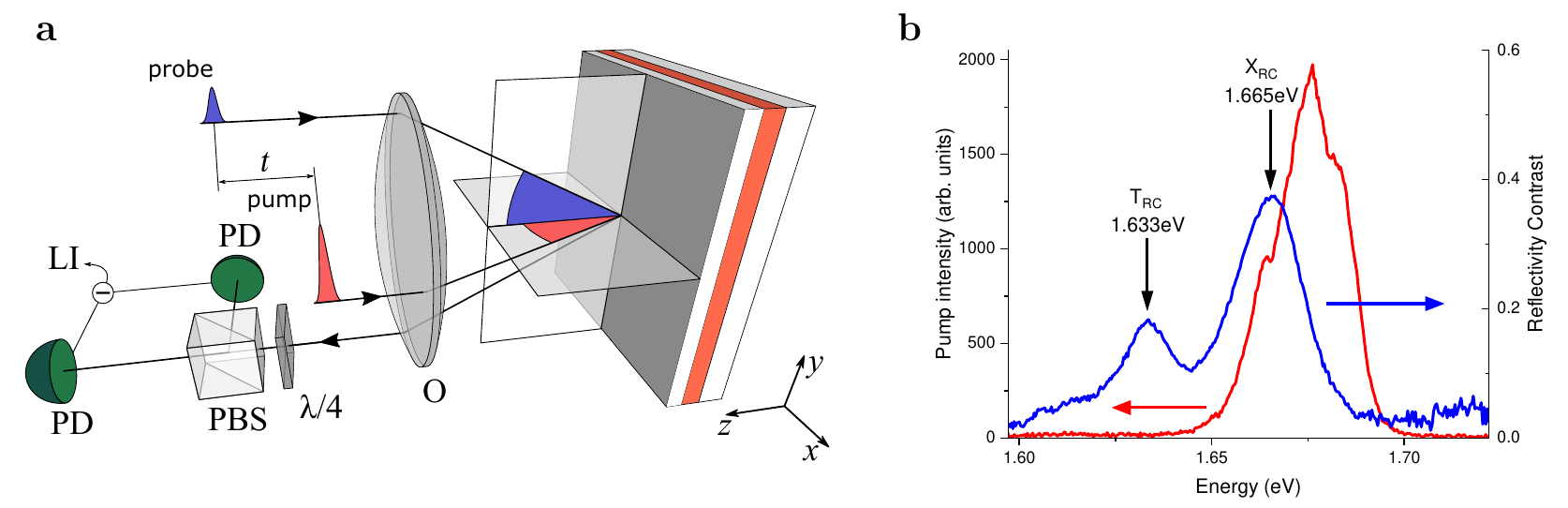}
	\end{center}
	\caption{(a) Illustration of the pump-probe system used for these measurements. O is the microscope objective, $\lambda/4$ the quarter-wave plate, PBS the polarizing beam splitter, PD are the balanced photodetectors and LI is the lock-in amplifier. (b) Comparison of the reflectance contrast ($\Delta R/R$) spectrum from Sample 2 with the pump spectrum, quasi-resonant with the neutral exciton transition. }
	\label{Dortmund_fig1}
\end{figure*}

We have \TU{studied spin dynamics using the time resolved pump-probe ellipticity technique} on a second MoSe$_2$ / EuS sample, which we call Sample 2. Reflectance contrast measurement of the monolayer reveals a spectrum similar to sample 1 presented in the main text, with two strong absorption features corresponding to the neutral exciton \TU{$X_{RC}$} (repulsive polaron) at higher energy and trion \TU{$T_{RC}$} (attractive polaron) at lower energy. 

For the time resolved pump-probe measurements (setup shown schematically in Fig.~\ref{Dortmund_fig1}a), \TU{we use laser pulses which are generated by a Ti:sapphire oscillator with a central wavelength of 790 nm, a spectral width of about 100~nm, and a repetition rate of 80~MHz. The laser beam travels through a pulse shaper and a compressor, which are used to compensate the pulse chirp acquired during propagation through the optical elements before reaching the sample. The resulting spectrum is centered around 765~nm (1.62~eV) with a full-width at half maximum (FWHM) of about 45 nm (100~meV). After that the laser beam is divided by a silica beam splitter into the pump and probe beams. The temporal delay $t$ between the pump and probe pulses is adjusted by a motorized mechanical delay line mounted in the pump beam path. The intensity of the pump beam is modulated using a mechanical chopper at the frequency of 2.1~kHz.}

\TU{In order to perform two-color pump-probe experiments each of the two beams is spectrally narrowed before focusing at the sample using interference bandpass filters. For pump pulses the center wavelength is set fixed at 740~nm (1.675~eV) with 10 nm bandwidth, leading to the photoexcitation of excitons in the $K$ valley of MoSe$_2$. The probe beam} has wavelength selectivity provided by a tunable bandpass filter \TU{with FWHM of 7~nm} in the excitation path, allowing a combination of time and spectrally resolved measurements. 
A comparison of the pump spectrum with the sample reflectance contrast spectrum (acquired with spectrally broad probe beam only, i.e. without using the tunable filter in the excitation path) is shown in Fig.~\ref{Dortmund_fig1}b. The duration of the pump and probe pulses are about 80 and 120~fs providing the overall time-resolution of about 150~fs.

\TU{The laser beams are focused at the sample surface into spots with diameters of about 20 $\mu$m using a single microscope objective (10$\times$ magnification with numerical aperture of 0.26). The pump power is set to 100 $\mu$W while the probe power varies from 1 to 10 $\mu$W depending on the central wavelength, i.e. the probe intensity is} at least $\sim10 \times$ weaker than the pump. \TU{The incidence angles for the pump and probe are 10$^\circ$ and 7$^\circ$ in the horizontal (zx) and vertical (zy) planes, respectively, as shown in Fig.~\ref{Dortmund_fig1}(a). The reflected beams are collected by the same objective. Then the probe beam is guided into the polarization bridge setup equipped with a polarizing beam splitter (PBS) and balanced photodetectors (PDs) [see Fig.~\ref{Dortmund_fig1}(a)]. The pump beam is circularly polarized, while the probe is linearly polarized. An additional quarter-wave plate is used before the PBS in order to measure the ellipticity signal, i.e. the difference between $\sigma^+$ and $\sigma-$ polarized components of the reflected probe beam. The PD signal is detected using a lock-in amplifier synchronized at the pump modulation frequency.} The resulting ellipticity signal reveals the time decay of the difference in $\sigma^+$ and $\sigma^-$ circularly polarized absorption from the sample, \TU{which is proportional to the pump induced spin polarization of photoexcited and resident carriers.}

\begin{figure*}[h]
	\begin{center}
		\includegraphics[width=1.0\textwidth]{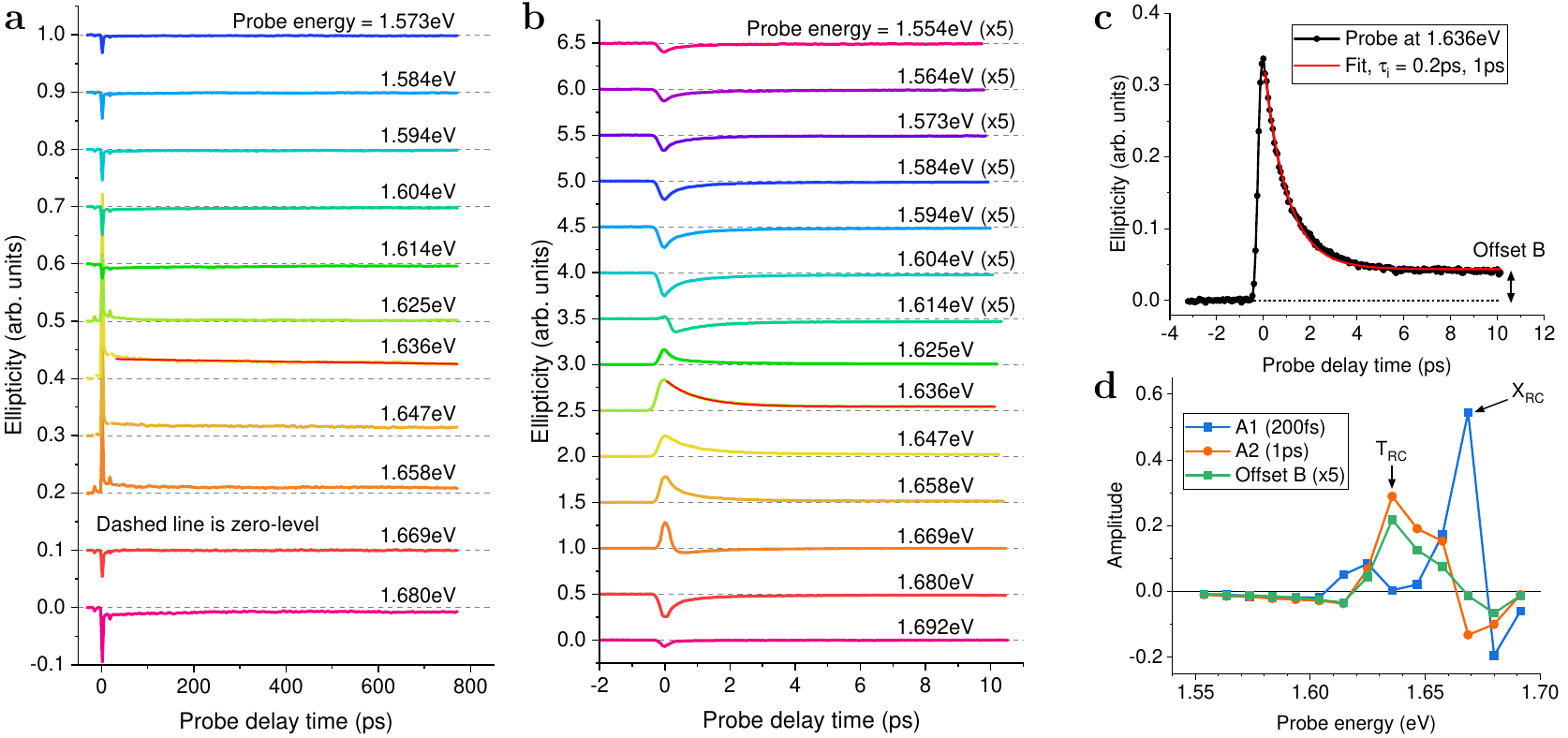}
	\end{center}
	\caption{(a,b) Transient ellipticity (quantifiying the difference between $\sigma^+$ and $\sigma^-$ reflectivity) over long (a) and short (b) timescales, for various probe energies. A very long decay component, fitted to $\sim3$ ns, can be seen in (a) when the probe is resonant with the trion state \TU{(1.636~eV)}, while on shorter timescales, the trion shows a biexponential decay (b). (c) Example of biexponential fitting to the transient ellipticity with the probe at \TU{1.636~eV}. The two timescales are 200~fs and 1~ps. The long ($\sim3$ ns) decay component can be seen as a clear offset from zero. (d) Amplitudes of the two components of the biexponential decay, and the offset / long decay versus probe energy. The 200 fs decay is strongest at the neutral exciton energy \TU{(1.665~eV)} while the 1 ps and 3 ns decays are strongest at the trion energy.  }
	\label{Dortmund_fig2}
\end{figure*}

Fig.~\ref{Dortmund_fig2} shows the results of the time resolved pump-probe spectroscopy. \TU{Each transient is normalized by the intensity of the probe beam for each particular probe photon energy $\hbar\omega_{\rm probe}$. Figure}~\ref{Dortmund_fig2}a shows the transient ellipticity on long (100s ps) timescales, in which a clear offset from zero can be seen when the probe is close to resonance with the trion state at \TU{1.636~eV}. This implies a long spin relaxation component, which we fit to be $\sim 3$ ns, and attribute to the spin relaxation of the free electrons constituting the 2DEG \cite{schwemmer2017long}. Fig.~\ref{Dortmund_fig2}b shows the measurement on a much shorter timescale (few ps), in which we observe a biexponential decay of the ellipticity. We fit these data with a biexponential curve superimposed on a constant offset $B$ from zero (see Fig.~\ref{Dortmund_fig2}c) \TU{$A_1\exp(-\frac{t}{\tau_{s1}})+A_2\exp(-\frac{t}{\tau_{s2}})+B$}, which accounts for the long-lived spin relaxation component seen in Fig.~\ref{Dortmund_fig2}a. This assumption is valid on the timescales considered here. \TU{From data fits at different energies $\hbar\omega_{\rm probe}$ it follows that there are two characteristic timescales with spin relaxation times $\tau_{s1}=200$~fs and $\tau_{s2}=1$~ps. The spectral dependence of $A_1$ and $A_2$} associated with the biexponential decay is extracted as a function of \TU{$\hbar\omega_{\rm probe}$} and plotted alongside the long-lived decay $B$ in Fig.~\ref{Dortmund_fig2}d. Here it can be seen that the maximum amplitude of \TU{$A_1(\hbar\omega_{\rm probe})$} occurs when the probe is resonant with the neutral exciton \TU{(1.665~eV)}, while \TU{$A_2(\hbar\omega_{\rm probe})$} and the offset \TU{$B(\hbar\omega_{\rm probe})$} both reach maximum values when the probe is resonant with the trion state \TU{(1.635~eV)}. This allows us to assign \TU{three contributions in the following way: $A_1$ with $\tau_{s1}=200$~fs} to neutral exciton spin relaxation, \TU{$A_2$ with $\tau_{s2}=1$~ps} to trion spin relaxation, and the offset value \TU{$B$ with decay constant of} $\sim3$~ns to the 2DEG spin relaxation \cite{schwemmer2017long}. This result corroborates our theoretical model presented in Supplementary Note 2 to explain the spin dynamics of the MoSe$_2$ microcavity system.

\clearpage

%
